\documentclass[aps,prl,twocolumn]{revtex4-1}
\usepackage{graphicx}
\usepackage{amsmath}
\usepackage{dcolumn}
\usepackage{bm}
\usepackage{float}
\usepackage{epstopdf}
\begin{document}

\preprint{}

\title{Comparing the anomalous Hall effect and the magneto-optical Kerr effect through antiferromagnetic phase transitions in Mn$_3$Sn}

\author{A. L. Balk, N. H. Sung, S. M. Thomas, P. F. S. Rosa, R. D. McDonald, J. D. Thompson, E. D. Bauer, F. Ronning, and S. A. Crooker}

\affiliation{Materials Physics and Applications Division, Los Alamos National Laboratory, Los Alamos, NM 87545, USA}

\begin{abstract}
In the non-collinear antiferromagnet Mn$_3$Sn, we compare simultaneous measurements of the anomalous Hall effect (AHE) and the magneto-optical Kerr effect (MOKE) through two magnetic phase transitions: the high-temperature paramagnetic/antiferromagnetic phase transition at the N\'eel temperature ($T_N \approx$420~K), and a lower-temperature incommensurate magnetic ordering at $T_1 \approx$270~K.  While both the AHE and MOKE are sensitive to the same underlying symmetries of the antiferromagnetic non-collinear spin order, we find that the transition temperatures measured by these two techniques unexpectedly differ by approximately 10~K. Moreover, the applied magnetic field at which the antiferromagnetic order reverses is significantly larger when measured by MOKE than when measured by AHE. These results point to a difference between the bulk and surface magnetic properties of Mn$_3$Sn.
\end{abstract}
\maketitle

Non-collinear antiferromagnets such as Mn$_3$Sn and Mn$_3$Ge have recently emerged as a fascinating class of materials that can exhibit a large anomalous Hall effect (AHE) despite having a negligibly small net magnetic moment \cite{Kubler, Nakatsuji, Nayak, Kiyohara, Nak2, XLi}. The AHE can arise in these and related antiferromagnetic (AF) materials when the underlying spin order not only breaks time-reversal symmetry but also lacks additional spatial symmetries that would otherwise force the AHE to vanish. Together with spin-orbit coupling, this can lead to a band exchange splitting and a non-zero value of the integrated Berry curvature over the occupied bands \cite{Chen, Suzuki, Nagaosa, Xiao, Zhang}, even in the absence of net magnetization. 

A related phenomenon that is also traditionally associated with the presence of a net magnetic moment is the magneto-optical Kerr effect (MOKE), wherein linearly-polarized light rotates and/or becomes elliptically polarized upon reflection from a material's surface.  Although MOKE is inherently a much more surface-sensitive probe than AHE, both phenomena result from off-diagonal components of the material's conductivity tensor, as discussed recently \cite{Feng} (e.g., $\sigma_{xz}(\omega)$ -- terms of this form generate currents that are transverse to applied electric fields). Such off-diagonal conductivity terms can in fact be non-zero in materials with specific non-collinear antiferromagnetic order, as shown recently \cite{Kubler, Nakatsuji, Nayak, Nak2, Chen, Suzuki, Zhang}. As such, anomalously large MOKE signals were also predicted in certain non-collinear antiferromagnets \cite{Feng}, and indeed they were very recently observed in Mn$_3$Sn by Higo \textit{et al.} \cite{Higo}. Both the AHE and MOKE are of practical interest as they can enable simple electrical and optical probes of non-collinear AF order, analogous to their widespread use to study ferromagnets. More fundamentally, both effects provide experimental tests for theoretical models \cite{Kubler, Chen, Suzuki, Nagaosa, Xiao, Zhang, Feng, Liu, Yang} that predict the influence of spin structure on measurable properties, based on underlying symmetry considerations.

Mn$_3$Sn has a hexagonal crystal structure ($P6_3 /mmc$ space group), with ``-AB-AB-" stacking of the planes along the [0001] direction.  Each plane contains a kagome lattice of Mn spins, as depicted in Fig. 1(a). Mn$_3$Sn exhibits a rich magnetic phase diagram, beginning (at high temperatures) with a paramagnetic-to-antiferromagnetic phase transition at its N\'eel temperature $T_N \approx 420~$K. Below $T_N$, neutron diffraction measurements \cite{Tomiyoshi, Nagamiya, Brown} indicate an in-plane inverse-triangular AF ordering of the Mn spins shown schematically in Fig. 1(a). This non-collinear AF state is characterized by nearly perfect compensation of the Mn spins within a unit cell, with only a very small residual in-plane magnetic moment of $\sim$0.003 $\mu_B$/Mn remaining. Furthermore, slightly Mn-deficient crystals also exhibit an additional first-order magnetic phase transition below room temperature at $T_1 \approx 270$~K, that is believed to reflect a change from a commensurate inverse-triangular magnetic order to an incommensurate spin structure that is helically modulated along the [0001] direction \cite{Zimmer, Kren, Ohmori, Cable}. This leads to a collapse of the residual net moment, and recent studies have also shown that the AHE also disappears below $T_1$ \cite{Li, Nakheon}, indicating a change in the underlying symmetry of the AF order. While these AF phase transitions in Mn$_3$Sn have been studied with neutron scattering and by electrical means, it is not yet known how they influence surface-sensitive MOKE signals, which to date have  been reported only near room temperature \cite{Higo}. 

Here, we perform simultaneous MOKE and AHE measurements of slightly Mn-deficient Mn$_{2.97}$Sn$_{1.03}$ (henceforth referred to as Mn$_3$Sn) as it is temperature-tuned through its AF phase transitions at $T_1$ and $T_N$. Between $T_1$ and $T_N$, both methods evince sizable signals due to the inverse-triangular AF order, as well as a large hysteresis in applied magnetic fields $B$ that arises from field-induced reversal of the AF order. However, the coercive field measured by MOKE ($\approx$120~mT) is over twice as large as that measured by AHE ($\approx$50~mT). Moreover, while both the MOKE and AHE signals vanish at low temperatures below $T_1$, the actual transition temperatures measured by the two techniques unexpectedly differ by about 10~K. We also observe $\sim$10~K difference in $T_N$. These results point to different magnetic behavior at the surface of Mn$_3$Sn as compared to the bulk.  

We study Mn$_3$Sn crystals grown by the molten metal self-flux method \cite{Nakheon, Canfield}. Samples were cut and mechanically polished to a mirror finish using 0.05 $\mu$m grit polishing paper. Laue diffraction confirmed that the surface prepared for MOKE studies is within 30 mrad of the (0001) crystal plane. As shown in the experimental schematic of Fig. 1(b), 25~$\mu$m diameter Pt wires were spot-welded on the (01$\bar{1}$0) face for AHE measurements. The sample was mounted between the poles of an electromagnet on a temperature-controlled stage with 100 mK stability. Magnetic fields $B$ were applied in the kagome plane, along the [01$\bar{1}$0] direction. All measurements were performed in a dry air environment.  We measured longitudinal MOKE, using 632.8 nm P-polarized light incident at 45$^\circ$ from the surface normal along the [01$\bar{1}$0] direction, as depicted.  The spot diameter on the sample is 5 $\mu$m and the Kerr rotation $\theta_K$ imparted on the reflected light is measured by balanced photodiodes.  Simultaneously, we measured the AHE by applying an ac current along the [0001] direction (perpendicular to the kagome planes) while detecting the Hall resistivity $\rho_H$ along the in-plane [2$\bar{1}\bar{1}$0] direction using standard lock-in techniques.  We note that prior studies have established that both the AHE and MOKE are very anisotropic in Mn$_3$Sn \cite{Nakatsuji, Nayak, Kiyohara, Higo, XLi}, with hysteretic signals vanishing when $B$ is applied along the [0001] direction (i.e., perpendicular to the kagome planes), and maximized when $B$ lies in the kagome planes.  The MOKE and AHE geometries that we use here are both chosen to be sensitive to the anomalous signals that arise from the inverse-triangular antiferromagnetic ordering of the Mn spins \cite{Nakatsuji, Nayak, Kiyohara, Higo}.

\begin{figure}[tbp]
\center
\includegraphics[width=.45\textwidth]{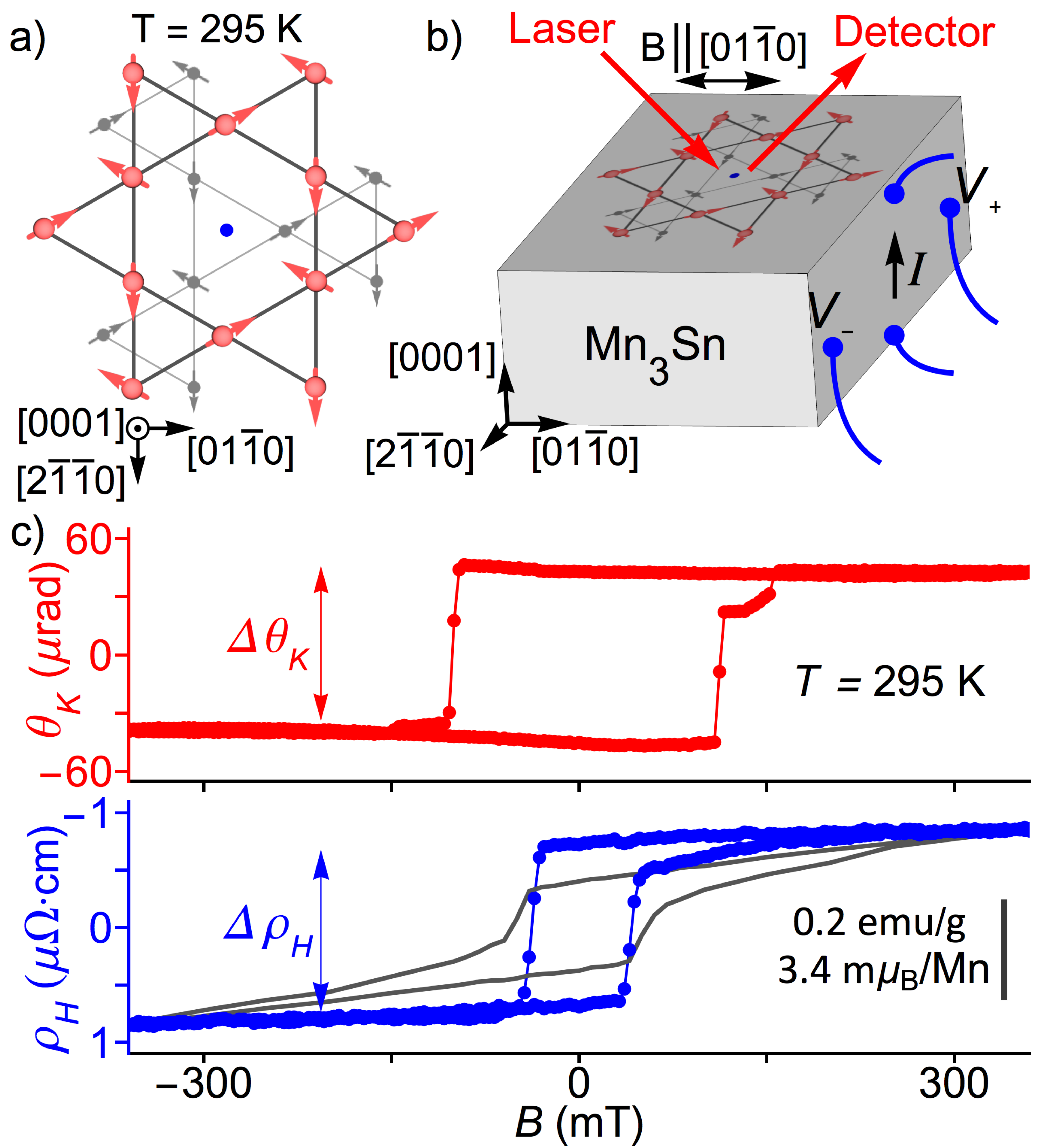}
\caption{(a) Non-collinear inverse-triangular AF spin structure of Mn$_{3}$Sn (at room temperature). Red arrows indicate Mn spins, the blue dot represents Sn. The crystal plane beneath the (0001) surface plane is depicted with reduced (grey) contrast. (b) Experimental setup. Longitudinal MOKE is measured on the (0001) surface while the AHE is sensed along [2$\bar{1}\bar{1}$0] direction using current along the [0001] direction. Magnetic fields $B$ are applied along [01$\bar 1$0]. The Mn$_3$Sn crystal dimensions are 2~mm $\times$ 2~mm $\times$ 1~mm.  (c) Simultaneous measurements of MOKE (red, top) and AHE (blue, bottom) at room temperature, versus $B$. Magnetic hysteresis is observed, showing transitions of magnitude $\Delta \theta_K$ and $\Delta \rho_H$. Note the very different AF switching (coercive) field. The grey loop is the bulk magnetization measured separately via SQUID magnetometry with $B$ along [01$\bar 1$0].}
\label{fig1} 
\end{figure}

Figure 1(c) shows both $\theta_K$ and $\rho_H$ measured in Mn$_3$Sn at room temperature ($T$=295~K) versus $B$. Both show large signals, with clear switching and magnetic hysteresis. As discussed above and as shown in previous experiments \cite{Nakatsuji, Nayak, Higo, Nakheon}, these large signals originate from the symmetry properties of the underlying inverse-triangular spin order. The ability to reverse the sense of this AF order (and therefore switch the sign of the AHE and MOKE signals) is due to the  residual net moment which, together with $B$, acts as a ``lever'' to invert the underlying AF magnetic structure. The amplitudes of the hysteresis loops, $\Delta \theta_K$ and $\Delta \rho_H$, are in agreement with recent studies \cite{Nakatsuji, Nayak, Higo, Nakheon}. However, the switching (coercive) field measured by MOKE is over twice that measured by AHE (120 mT vs. 50 mT). This marked contrast provides a first indication that the surface and bulk magnetic behavior of Mn$_3$Sn is not the same. For reference, the grey curve in Fig. 1(c) shows the bulk magnetization of this sample acquired by SQUID magnetometry, where the switching of the AF order is revealed by the concomitant switching of the small 0.003 $\mu_B$/Mn residual moment. The coercive field coincides with that measured by the AHE, consistent with the expectation that the AHE is sensitive to bulk magnetic properties. Subtle differences in the shape of the AHE and magnetization hysteresis loops have been discussed recently in the context of real-space Berry curvature due to AF domain walls \cite{XLi}.

We now compare MOKE and AHE signals as the Mn$_3$Sn sample is cooled below room temperature and through its phase transition at $T_1$. Figures 2(a-c) show both measurements at selected temperatures: At 278~K, both continue to exhibit large signals and robust magnetic hysteresis. However, at 267~K the two signals differ dramatically -- the AHE continues to show a substantial signal and clear magnetic hysteresis, while in contrast MOKE shows no signal (and no hysteresis). We emphasize that these measurements were performed simultaneously, indicating that the bulk and the surface of the sample exhibit very different magnetic behavior at this temperature. Finally, at 261~K both methods show no signal, indicating that both the bulk and the surface have transitioned to the low-$T$ incommensurate AF phase. 

\begin{figure}[tbp]
\center
\includegraphics[width=.40\textwidth]{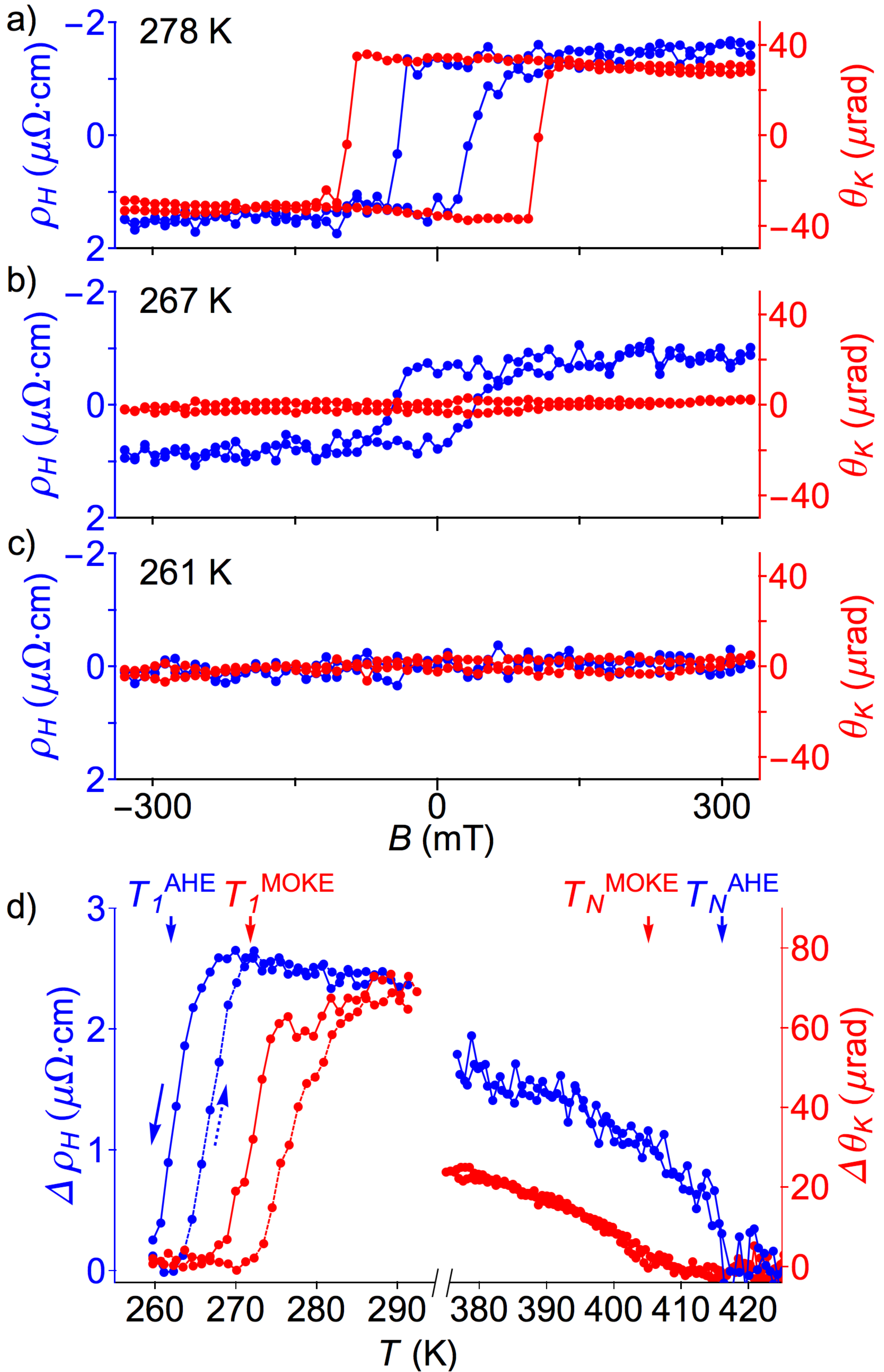}
\caption{(a-c) Simultaneous measurements of AHE and MOKE versus $B$, at temperatures 278~K, 267~K, and 261~K. Note that at 267~K the two measurements show very different behavior, indicating a marked difference between the bulk and surface magnetic properties. (d) Temperature dependence of the amplitudes of the magnetic hysteresis loops, as measured by AHE (blue) and by MOKE (red). Both $\Delta\rho_H$ and $\Delta\theta_K$ reveal the first-order AF phase transition near $\approx$270~K; however the measured transition temperature $T_1^{\rm MOKE}$ is approximately 10~K higher than $T_1^{\rm AHE}$. Moreover, the measured N\'eel temperature $T_N^{\rm MOKE}$ is $\sim$10~K lower than $T_N^{\rm AHE}$.}
\label{fig2}
\end{figure}

To explore this difference in more detail, both $\rho_H (B)$ and $\theta_K (B)$ are measured continuously as the temperature is ramped from 300 K down to 260 K, and then back up to 300 K (at 0.05 K/s). The amplitudes of the magnetic hysteresis loops, $\Delta\rho_H$ and $\Delta\theta_K$, are shown in Fig. 2(d). The AHE reveals a first-order phase transition in the bulk of the sample at $T_{1}^{\rm AHE}$ = 265~K (with $\sim$5 K of thermal hysteresis), consistent with recent work \cite{Nakheon}. However, the MOKE data reveal a rather different transition temperature, $T_{1}^{\rm MOKE}$ = 275~K, at the surface of the sample. This 10~K difference in $T_{1}$ is much larger than any experimental uncertainty in temperature, and was confirmed in multiple temperature sweeps with the probe laser positioned at different locations on the (0001) surface plane.  We note that this difference cannot arise from artifacts due to thermal gradients in the sample: the top surface of the sample on which MOKE is detected must be, if anything, slightly warmer than the bulk of the crystal and the sample stage (because $T_1$ and the sample stage are below the temperature of the surrounding dry-air environment), which would lead to a slightly \textit{lower} apparent $T_{1}^{\rm MOKE}$ in the experiment -- opposite to what is observed in Fig. 2(d).

Similarly, we also investigate whether MOKE and AHE show different N\'eel temperatures $T_N$ at the high-temperature antiferromagnetic-paramagnetic phase transition. Fig. 2(d) shows that both $\Delta\theta_K$ and $\Delta\rho_H$ vanish as the temperature is increased, indicating a traversal of $T_N$. Upon subsequent cooling, these curves are re-traced without discernible thermal hysteresis. However, once again the data reveal $\approx$10 K difference between the transition temperatures. In this case, however, $T_N^{\rm MOKE}$ is lower than $T_N^{\rm AHE}$. As before, this difference cannot be due to thermal artifacts between the sample and the surrounding environment: a cooler temperature at the surface than in the bulk would result in a higher apparent $T_N^{\rm MOKE}$, in contrast to observation. These measurements therefore indicate that the surface and bulk of Mn$_3$Sn undergo AF phase transitions at different temperatures.

Finally we explore in more detail the large factor-of-two disparity between the AF switching field (\textit{i.e.}, the coercive field $\mu_0H_c$) measured by MOKE and by AHE, that was shown earlier in Figs. 1 and 2. To check whether this large difference could be due to local extrinsic pinning from isolated defects at the sample surface, we measure $\theta_K (B)$ at fifty random locations on the (0001) surface plane, each separated by $>50~\mu$m. Six representative hysteresis loops are shown in Fig. 3(a). There is scatter in $\mu_0H_c$, revealing some influence of extrinsic pinning forces. However, a histogram of all measured $\mu_0H_c$ values [Fig. 3(b)]reveals a mean value of 120~mT, with only $\pm$20 mT variation that is far smaller than the $\sim$70~mT difference between $\mu_0H_c$ measured by MOKE and by AHE.

Various mechanisms could account for the unexpected differences between the values of $T_{1}$, $T_N$, and $\mu_0H_c$ that are measured in the bulk of Mn$_3$Sn (by AHE) and at the surface (by MOKE). We estimate the penetration depth of the 632.8~nm probe light in Mn$_3$Sn to be of order 20 nm (based on carrier densities reported in \cite{Li}), which significantly exceeds the $\sim$5~nm lengthscale of the helical modulation that is believed to exist in the low-temperature incommensurate AF phase below $T_1$ \cite{Tomiyoshi, Nagamiya, Brown}. Since no indication of smaller bulk-like coercive fields are observed in the MOKE data, nor are the AF transitions at $T_1$ and $T_N$ noticeably less sharp that those measured by AHE, the different surface magnetic properties of Mn$_3$Sn likely extend within the sample on at least this length scale. Surface oxidation could influence the magnetism detected by MOKE. To test this we re-polished the (0001) surface plane and then (within 15 minutes) continuously measured $\theta_K (B)$ hysteresis loops over several hours in ambient conditions. We did not observe any change in $\mu_0H_c$, arguing against slow surface oxidation as a cause for the different magnetic behavior. It is also possible that the surface preparation itself causes local disorder which increases $\mu_0 H_c$ and changes $T_1$ and $T_N$ due to increased pinning forces, although we note that good crystal quality at the (0001) surface was confirmed by clean Laue diffraction signals. Finally, preliminary studies of other Mn$_3$Sn samples under applied uniaxial strain \cite{Sean} allow us to extrapolate and estimate that over 3\% strain would be necessary to account for the observed 10 K change in $T_1$.  This estimated value, while large, is in fact comparable to surface strains induced by mechanical polishing that have been measured in other materials \cite{Hang, Shen}. 

\begin{figure}[tbp]
\center
\includegraphics[width=.40\textwidth]{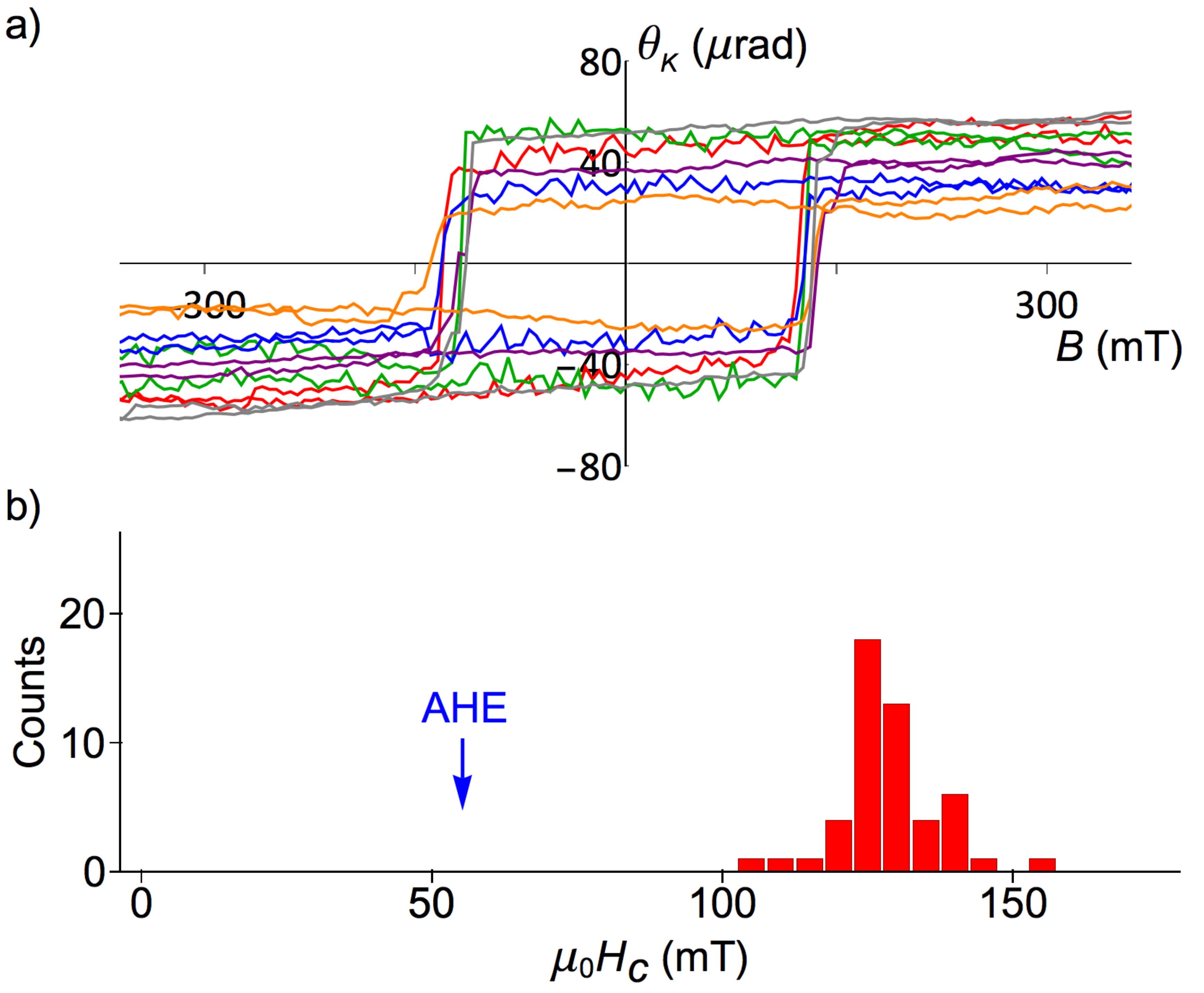}
\caption{(a) The plot shows a few of the fifty MOKE hysteresis loops ($\theta_K$ vs. $B$) that were measured at different locations on the Mn$_3$Sn (0001) surface plane. $T$=295~K. The locations were separated by $>$50~$\mu$m, and reveal some variation in the AF switching (coercive) field $\mu_0H_c$ at the surface of Mn$_3$Sn. (b) Histogram of $\mu_0H_c$ values measured with MOKE. The much smaller switching field measured in the Mn$_3$Sn bulk by the AHE is indicated by the blue arrow.}
\label{fig3}
\end{figure}

We note that differences between surface and bulk magnetism have been observed in other AFs such as NiO, GdIn$_3$, and UO$_2$ \cite{Marynowski, Malachias, Watson, Pleimling} where different exchange forces,  stoichiometry, or disorder at the sample surface can lead to phase transitions with different temperatures as compared to the bulk. In Mn$_3$Sn, a smaller $T_N$ at the surface is consistent with decreased AF exchange interactions at the surface. This is at least in line with neutron scattering results \cite{Cable}, which show significant inter-plane exchange interactions along the [0001] direction in Mn$_3$Sn.  Whether this can also account for the \textit{larger} value of $T_1$ at the surface is not yet clear. It is also worth noting that the slightly Mn-rich crystals studied recently by Higo \cite{Higo} do not appear to exhibit substantially different switching fields when studied by MOKE and by AHE. The recent availability of Mn$_3$Sn thin films \cite{Markou} should allow further studies of these phenomena and closer comparisons of various experimental techniques to probe non-collinear AF order.

In summary, we have shown that MOKE measurements can be used to probe temperature-dependent phase transitions in non-collinear antiferromagnets such as Mn$_3$Sn. Similar to the AHE, MOKE is directly sensitive to the symmetry properties of the underlying magnetic order, providing a facile means to study AF order in this class of materials. Unexpectedly, simultaneous MOKE and AHE studies reveal different transition temperatures $T_1$ and $T_N$, as well as significantly different AF switching (coercive) fields in the inverse-triangular AF phase. These results point to different surface and bulk magnetic properties, which may be relevant for potential applications using Mn$_3$Sn or other non-collinear antiferromagnets.

We thank C. Batista for helpful discussions. This work was supported by the Los Alamos LDRD program. The MOKE and AHE studies were performed at the National High Magnetic Field Laboratory (NHMFL) at Los Alamos, which is supported by NSF DMR-1644779, the State of Florida, and the US Department of Energy.

\end{document}